\documentclass[10pt,twocolumn]{article}
\usepackage{amssymb}
\usepackage{marvosym}
\usepackage{graphicx}
\usepackage{epstopdf}

\title{2D Direction Of Arrival Estimation with Modified Propagator}
\author{Y.Khmou, S.Safi\\
        Department of Mathematics and Informatics,\\
        polydiscplinary faculty, Beni Mellal, Morocco.}

\begin{document}
\maketitle
\begin{abstract}
In this paper, a fast algorithm for the Direction Of Arrival (DOA) estimation of radiating sources, based on partial covariance matrix and without eigendecomposition of incoming signals is extended to two dimensional problem of joint azimuth and elevation estimation angles using Uniform Circular Array (UCA) in case of non coherent narrowband signals. Simulation results are presented  with both Aditive White Gaussian Noise (AWGN) and real symmetric Toeplitz noise.
\end{abstract}
\section{INTRODUCTION}
Estimating the Direction of Arrivals (DOA) of impinging electromagnetic or acoustic waves on antenna array   plays a crucial role in many physics and engineering fields such as sonar,radar, astrophysics,Electronic Surveillance Measure (MSE),submarine acoustics, geodesic location , GSM positionnging,  geophysics, and so on.\\ \\
Extensive research has been made in  the last two decades leading to an efficient improvement in DOA estimation but every method has its own advantages and disadventages in terms of the nature of environement such as the jamming,reflection, thermal noise, degradation of the antenna elements  in the array ,the radiation pattern of the array ,near field scattering and mutual coupling .\\ \\
A subspace based methods provide high resolution estimation [1],[2], Maximum Likelihood Estimation (ML) and Estimation of Signal Parameters via Rotational Invariance Technique (EPRIT) [3] were proposed in various array geometries such as a Uniform Linear Array (ULA), Uniform Rectangular Array (URA). Howevere the subspace based techniques require extensive computation of the eigendecomposition of covariance matrix of the array output data in order to obtain the sets of signal and noise subspaces, which makes the use of these techniques limited in case of large number of array sensors.\\ \\
For this reason, a fast DOA estimation methods were proposed [4], [5], [6] without eigendecomposition ,  the propagation method [7],[8]  uses the whole covariance matrix of the output data to obtain the propagation operator but it can be degraded in non uniform colored noise [9] . \\
To overcome this issue, a modified propoagation method (MPM) [10] were proposed with Uniform Linear Array (ULA) using only partial covariance matrix which makes it lower computationally than the (PM) method [7] and better performing with non Gaussian noise .\\ \\
In this paper, we present  an extension (MPM) method [10] to joint two dimensional azimuth and elevation estimation angles using Uniform Circular Array (UCA) with non moving narrowband radiating sources in both complex Additive White Gaussian Noise (AWGN) and real symmetric Toeplitz noise.
\section{ DATA MODEL}
We consider uniform cicrular array (UCA), consisting of N isotropic and identical antenna elements distributed uniformely over a circle with radius $R=\lambda/2$ with $\lambda$ is the wave length of the impinging electromagnetic waves . The phase azimuth angle of the $n^{th}$ element is $\phi_{n}={{2\pi n}\over{N}}$ with $n=1,2,...,N$ .  P narrowband source waves with same carrier frquency $f$ are impinging on the UCA with directions $(\theta_{1},\phi_{1}),(\theta_{2},\phi_{2}),...,(\theta_{P},\phi_{P}) $
The received signals can be written in  the following form :
\begin{equation}
X(t)=AS(t)+N(t)
\end{equation}
Where : $X(t)=[X_{1}(t),X_{2}(t),...,X_{N}(t)]^T$ is the $t^{th}$ snapshot of the received signal,
 $A=[a(\theta_{1},\phi_{1}),a(\theta_{2},\phi_{2}),...,a(\theta_{P},\phi_{P})]$
is the NxP array response with $a(\theta_{m},\phi_{m})=[a_{1}(\theta_{m},\phi_{m}),a_{2}(\theta_{m},\phi_{m}),...a_{N}(\theta_{m},\phi_{m})]^T$
is the steering vector from direction $(\theta_{m},\phi_{m})$ where $m=1,2,...,P$, $(\theta_m,\phi_m)$ are the elevation and  azimuth of the $m^{th}$ signal respectively.$S(t)=[s_{1}(t),s_{2}(t),...,s_{P}(t)]^T$ is the source waveform vector and $N(t)=[n_{1}(t),n_{2}(t),...,n_{N}(t)]^T$ is the $t^{th}$ snapshot of either  zero mean stationary complex additive white Gaussian noise (AWGN) or non uniform spatially and temporally complex colored noise . The steering vector a can be written as the following :\\
$a_{m}(\theta,\phi)=$
\begin{equation}
e^{2\pi j{R \over \lambda}\left ( \sin({\theta})\cos({\phi})\cos({{2\pi m}\over N})+sin({\theta})\sin({\phi})\sin({{2\pi m} \over N}) \right ) }
\end{equation}
Using the goniometric identity $\cos(a \pm b)=\cos(a)\cos(b) \mp \sin(a)\sin(b)$ the steering vector becomes :
\begin{equation}
a_{m}(\theta,\phi)=e^{2\pi j {R\over \lambda}\sin(\theta ) \cos({{2\pi m} \over N}-\phi)}
\end{equation}
In case of (AWGN) noise the covariance matrix can be written as :
\begin{equation}
R_{xx}=E \{ X(t) {X(t)^H} \} = AR_{ss}A^{H}+\sigma_{N}^{2}I
\end{equation}
Where $R_{ss}$ is the covariance matrix of the incident signals , H denotes the hermitian transpose and $I$ is NxN identity matrix .\\
The angles of arrivals (AOA) can be extracted using $R_{xx}$ when the the following assumptions are respected: 
(1) The number of sources is known a priori and the number of antenna elements in the UCA satisfies : $N \geqslant 2P+2$.\\
(2) The P steering vectors are linearly independent and the P signal sources are statistically uncorrelated.\\
(3) The radiating sources are located in the Fraunhofer field (far field) where the distance of the $m^{th}$ source satisfies $d_{m} \geqslant {2 D^{2} \over \lambda} $ where D is the maximal dimension of the UCA elements .
\section{THEORY OF THE MODIFIED PROPAGATOR }
the Propagator Method (PM) $[7],[8]$  is computationally low because it does not need eigendecomposition of the covariance matrix , but it uses the whole of it, to obtain the propagation operator. When the noise is not stationary  ergodic Gaussian, the PM can be degraded $[9]$, to overcome the problem ,we make an extension of  the  modified version (MPM) $[10]$ on Uniform Circular Array (UCA) as the following :\\
The array response matrix in equation (3) can be partitionned as the following [10] :
\begin{equation}
A(\theta,\phi)=[A_{1}^{T}(\theta,\phi),A_{2}^{T}(\theta,\phi),A_{3}^{T}(\theta,\phi)]^T
\end{equation}
The matrices $A_1$ and $A_2$ have dimensions $P\times P$ where $A_3$ is $(N-2P)\times P$. Based on the equation (1) and the above partition, we can extract the following partial cross correlation matrices from  $R_{xx}$ : \\
\begin{equation}
R_{12}=E \{ X_{t}[1:P,:] X_{t}^{H}[P+1:2P,:] \} = A_{1}R_{ss}A_{2}^{H}
\end{equation}
\begin{equation}
R_{31}=E \{ X_{t}[2P+1:N,:] X_{t}^{H}[1:P,:] \} = A_{3}R_{ss}A_{1}^{H}
\end{equation}
\begin{equation}
R_{32}=E \{ X_{t}[2P+1:N,:] X_{t}^{H}[P+1:2P,:] \} = A_{3}R_{ss}A_{2}^{H}
\end{equation}
The matrices $A_{i}$, $i=(1,2)$ and Rss are full rank according the assumption 2, therefore we can compute the matrix $A_{3}$ with two different ways :
\begin{equation}
R_{32}R_{12}^{-1}A_{1}=A_{3}R_{ss}A_{2} A_{2}^{-H}R_{ss}^{-1}A_{1}^{-1}A_{1}=A_{3}
\end{equation}
\begin{equation}
R_{31}R_{21}^{-1}A_{2}=A_{3}R_{ss}A_{1} A_{1}^{-H}R_{ss}^{-1}A_{2}^{-1}A_{2}=A_{3}
\end{equation}
Combining equations (14) and (15) gives :
\begin{equation}
R_{32}R_{12}^{-1}A_{1}+R_{31}R_{21}^{-1}A_{2}=2A_{3}
\end{equation}
Augmenting the left hand side of the equation (16) yields to :
\begin{equation}
\left [ R_{32} R_{12}^{-1} \hspace{0.5cm} R_{31} R_{21}^{-1} \hspace{0.5cm}  -2I_{N-2P}      \right ] A=0
\end{equation}
which is equivalent to :
\begin{equation}
Q^{H}A=0
\end{equation}
When the P signals are registered with corresponding directions $(\theta_{k},\phi_{k})$
the equation (18) holds :
\begin{equation}
Q^{H}a(\theta_{k},\phi_{k})=0  \hspace{2cm}  k=1,2,...P
\end{equation}
Finally when scanning the azimuth and elevation angles in the ranges $[0,2\pi]$ $[0,{\pi \over {2}}]$ respectively, the two dimensional spatial spectrum has peaks when encountring the true signals directions :
\begin{equation}
f(\theta,\phi)={1 \over{{\Vert  Q^{H} a(\theta,\phi)       \Vert}^2}}
\end{equation} 
Where the steering vector a is defined by :
\begin{equation}
a(\theta_{i},\phi_{i})= \left [e^{z_{i}\cos(\phi_{i})}, e^{z_{i} \cos(\phi_{i}-{2\pi \over N})},...,e^{z_{i} \cos(\phi_{i}-{2\pi(N-1) \over N})} \right ]
\end{equation}
where $z_{i}=2\pi j {R \over \lambda} \sin(\theta_{i})$, $(\theta_{i},\phi_{i})$ are the $i^{th}$ scanned angles with $\theta_{i} \in \left [ 0:{1 \over f_{\theta}}: {\pi \over 2} \right ]$,$\phi_{i} \in \left [ 0:{1 \over f_{\phi}}: 2\pi \right ]$ , $(f_{\theta},f_{\phi})$ are the desired spatial sampling frequencies .\\
\section{RESULTS AND DISCUSSION}
We consider a Uniform Circular Array (UCA) consisting of $N=14$ identical dipoles elements having the  same exitation and zero phase, with operation frequency $f=900MHz$, the radius of the UCA is $R=38cm$ with interlement spacing $d=16cm$ as illustrated in the figure 1.\\
Three non coherent and almost equipowered narrowband sources are impinging on the array from $(\theta_{1}=15^{\circ},\phi_{1}=20^{\circ})$,$(\theta_{2}=30^{\circ},\phi_{2}=44^{\circ})$ and $(\theta_{3}=66^{\circ},\phi_{3}=69^{\circ})$ respectively with $K=100$ snapshots .
\begin{figure}[!h]
\includegraphics[width=3.75in]{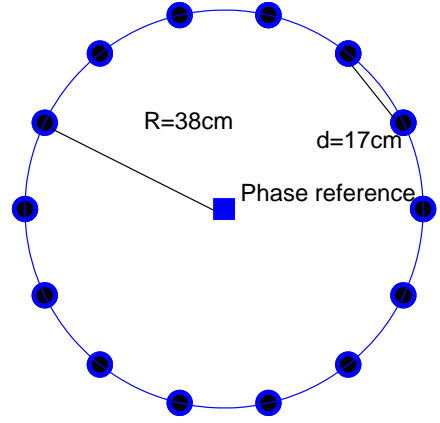}
\caption{UCA, $d=\lambda/2$.}
\end{figure}
The figure 1 represents the average of $L=50$ Monte Carlo simulation trials when the $SNR=10dB$.
\begin{figure}[!h]
\includegraphics[width=9cm]{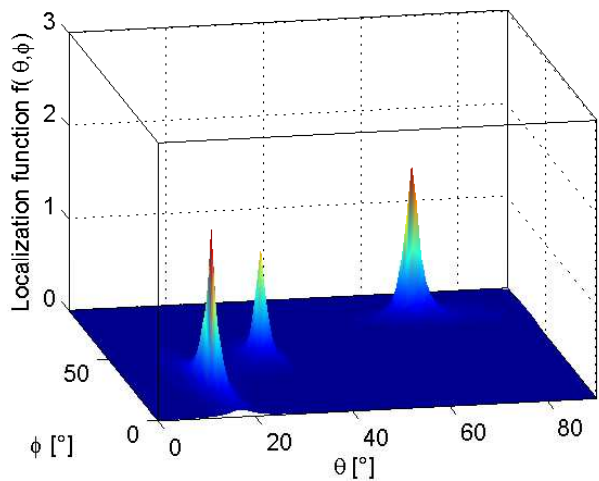}
\caption{Modified Propagator, $SNR=10dB$.}
\label{DOA_awgn}
\end{figure}
In the second simulation, we evaluated the (MPM) in case of real symmetric Toeplitz noise where the noise covariance is given by :
\begin{equation}
Rn=Toeplitz([1,0.93,0.86,...,0.1]) 
\end{equation}
\begin{figure}[!h]
\includegraphics[width=9cm]{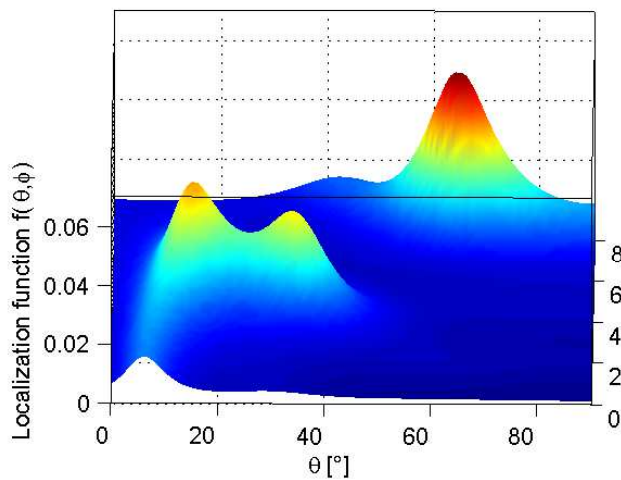}
\caption{Modified Propagator in real symmetric toeplitz noise}
\end{figure}
The MPM method can be succeful in the presence of non uniform noise .
The advantage of the modified propagator is complexity[10], while the high resolution MUSIC algorithm involves $N^{2}K$ for computing the covariance matrix $R_{xx}$ and $O(N^3)$ for its Eigen-Value Decomposition (EVD), and the number of floating points the  (PM) method takes for computing the propagator Q is $NPK+O(P^{3})$
, the MPM algorithm only involves $P(N-P)K+O(P^{3})$.\\ 
If the sources are correlated or coherent to each other, the algorithm will fail to detect the DOAs, to get the same result one needs to use preprocessing techniques on the covariance matrix such as the Forward Backward Averaging (FBA) or the spatial smoothing.
\section{Conclusion}
A Modified Propagator Method (MPM) for  DOA estimation has been extended to two dimensional azimuth and elevation angles using Uniform Circular Array (UCA), the partial cross correlation matrices are used to compute the propagation operator using off diagonals of the output cross correlation matrix, hence the MPM is suitable to the case of non uniform noise which is confirmed by simulation  in the presence of real symmetric Toeplitz noise .

\end{document}